\documentclass[letterpaper,10pt]{article} 

\usepackage{osameet3} 

\usepackage{amsmath,amssymb}
\usepackage[colorlinks=true,bookmarks=false,citecolor=blue,urlcolor=blue]{hyperref} 
\usepackage[nolist]{acronym}
\usepackage{caption}
\usepackage{subcaption}
\usepackage{float}
\usepackage{multirow}
\usepackage[export]{adjustbox}

\makeatletter
\newcommand\Headnote{\parbox{20cm}{This work has been published in Optical Fiber Communication Conference (OFC) 2020}}
\newcommand\Footnote{\parbox{16cm}{\hrulefill \\ \copyright~ 2020 Optical Society of America. One print or electronic copy may be made for personal use only. Systematic reproduction and distribution, duplication of any material in this paper for a fee or for commercial purposes, or modifications of the content of this paper are prohibited.}}

\let\Title\@title
\def\ps@mystyle{%
    \setlength{\voffset}{-0.5in}
    \setlength{\headsep}{0.5in}
    \def\@oddfoot{\hfill\Footnote\hfill}%
    \let\@evenfoot\@empty%
    \def\@evenhead{\hfill\Headnote\hfill}%
    \def\@oddhead{\hfill\Headnote\hfill}%
    \let\@mkboth\markboth}
\makeatother
\begin{document}

\title{PON Virtualisation with EAST-WEST Communications for Low-Latency Converged Multi-Access Edge Computing (MEC)}

\vspace{-0.1in}
\vspace{-4mm}
\author{Sandip Das, Marco Ruffini}
\vspace{-0.02in}
\address{CONNECT Center, Trinity College Dublin}
\vspace{-0.02in}
\email{dassa@tcd.ie, marco.ruffini@tcd.ie}

\copyrightyear{2020}

\vspace{-5mm}
\begin{abstract}
We propose a virtual-PON based Mobile Fronthaul (MFH) architecture that allows direct communications between edge points (enabling EAST-WEST communication). Dynamic slicing improves service multiplexing while supporting ultra-low latency under 100$\mu$s between cells and MEC nodes.
\end{abstract}
\vspace{-1mm}
\vspace{-5mm}
\begin{acronym}
\acro{QoS}{Quality of Service}
\acro{C-RAN}{Cloud Radio Access Networks}
\acro{FBG}{Fibre Bragg Grating} 
\acro{MFH}{Mobile Fronthaul}
\acro{RU}{Radio Unit}
\acro{BBU}{Baseband Unit}
\acro{DU}{Distributed Unit}
\acro{CU}{Central Unit}
\acro{PON}{Passive Optical Network}
\acro{vPON}{virtual-PON}
\acro{ODN}{Optical Distribution Network}
\acro{TWDM}{Time-Wavelength Division Multiplexing}
\acro{DBA}{Dynamic Bandwidth Allocation}
\acro{MEC}{Multi Access Edge Computing}
\acro{CO}{Central Office}
\acro{OLT}{Optical Line Terminal}
\acro{ONU}{Optical Networking Unit}
\acro{PLOAM}{Physical Layer Operation and Maintenance}
\acro{eCPRI}{evolved Common Public Radio Interface}
\acro{BS}{Base Station}
\acro{TDMA}{Time Division Multiple Access}
\acro{TTI}{Transmit Time Interval}
\acro{VRF}{Variable Rate Fronthaul}
\acro{vPON}{Virtualized PON}
\acro{UE}{User Equipment}
\acro{LLS}{Low Layer Split}
\acro{WLB}{Wavelength Loop Back}
\acro{WPF}{Wavelength Pass Filter}
\end{acronym}
\vspace{-1mm}
\section{Introduction}\vspace{-1mm}
    In the 5G era of telecommunications, characterised by diversified service requirements, low-latency and deterministic \ac{QoS} are two most crucial network requirements. With the drive towards \ac{C-RAN} to support these requirements, building a reliable and low-cost fronthaul transport network becomes a significant issue. For example, a \ac{C-RAN} that uses eCPRI (the evolved CPRI standard for 5G networks) has a very stringent latency requirement of $\approx$ 100$\mu s,$ for any split that is below the MAC layer. 
    In order to achieve this low transport network latency, the transmission distance between the cell site (\ac{RU}) and the first processing site (\ac{DU}) is made shorter by deploying some limited capacity cloud processing resources (called \ac{MEC}) close to the cell-sites.
    
    As 5G networks will bring a progressive densification of mobile cells, the cost of the optical transport network will soar to unsustainable levels, if cells are connected through dedicated point-to-point fibre. In addition, point-to-point solutions provide little flexibility in redirecting \ac{RU} connectivity between edge cloud nodes during migration events. 
    For this reason, \ac{PON}, typically used for providing broadband to residential and small business users, is being considered as a possible cost-effective solution to support optical \ac{MFH} transport, as it can use an already deployed \ac{ODN} to provide fronthaul transport for \acp{RU} along with serving residential users. 
    However, achieving low latency is a major challenge for \acp{PON} in the upstream direction, because of its \ac{TDMA} nature of operation. 
    A solution was proposed in \cite{M-DBA} and recently standardised by ITU-T \cite{G.989.3Am1}, called cooperative DBA, which adopts a mechanism where \ac{UE} scheduling information is passed directly from the \ac{DU} to the \ac{OLT}.  This bypasses the report/grant mechanism, thus enabling low \ac{MFH} transport latency. 
    
    While this workaround solves the latency issue when \acp{ONU} are connected to a given \ac{OLT} in a \ac{CO}, where also the \ac{DU} and \ac{CU} processing are located, it does not provide a solution for edge cloud environments, where the \ac{DU} processing might migrate across edge nodes, for example for load balancing or to support low latency also at the application level.
    
    In this work, we propose a novel architecture, based on \ac{PON} virtualisation, which also enables EAST-WEST \ac{PON} communication. This enables end points, where typically only end users or \acp{RU} would be located, to also host edge nodes. 
    In this way, for example, \acp{RU} can dynamically redirect their connection from \acp{DU}/\acp{CU} located at the central offices (i.e., at the source of the PON tree) towards ones located at the edge (i.e., at the leaves of the \ac{PON} tree). This largely improves the statistical multiplexing ability of the PON to support low-latency services.
    While the concept of dynamic 
    \ac{vPON} was also proposed in \cite{vPON}, there the authors restrict the location of edge nodes to the splitter nodes and consider dynamic offloading only between edge nodes and \ac{CO}. 
    
Our work introduces instead the ability to create virtual PONs across a mix of \ac{CO} and edge nodes, which can be located anywhere in the \ac{PON}. 
    In addition, we propose a novel \ac{CO}-assisted dynamic vPON slice formation mechanism for offloading ONUs between edge \acp{OLT}, to provide ultra-low end-to-end transport latency for \ac{MFH}.

    
\vspace{-1.5mm}
\section{Proposed \ac{vPON} architecture for MEC support through EAST-WEST communication} 
\begin{figure}[htbp]
    \begin{subfigure}{0.50\textwidth}
        \frame{\includegraphics[width=\textwidth,left]{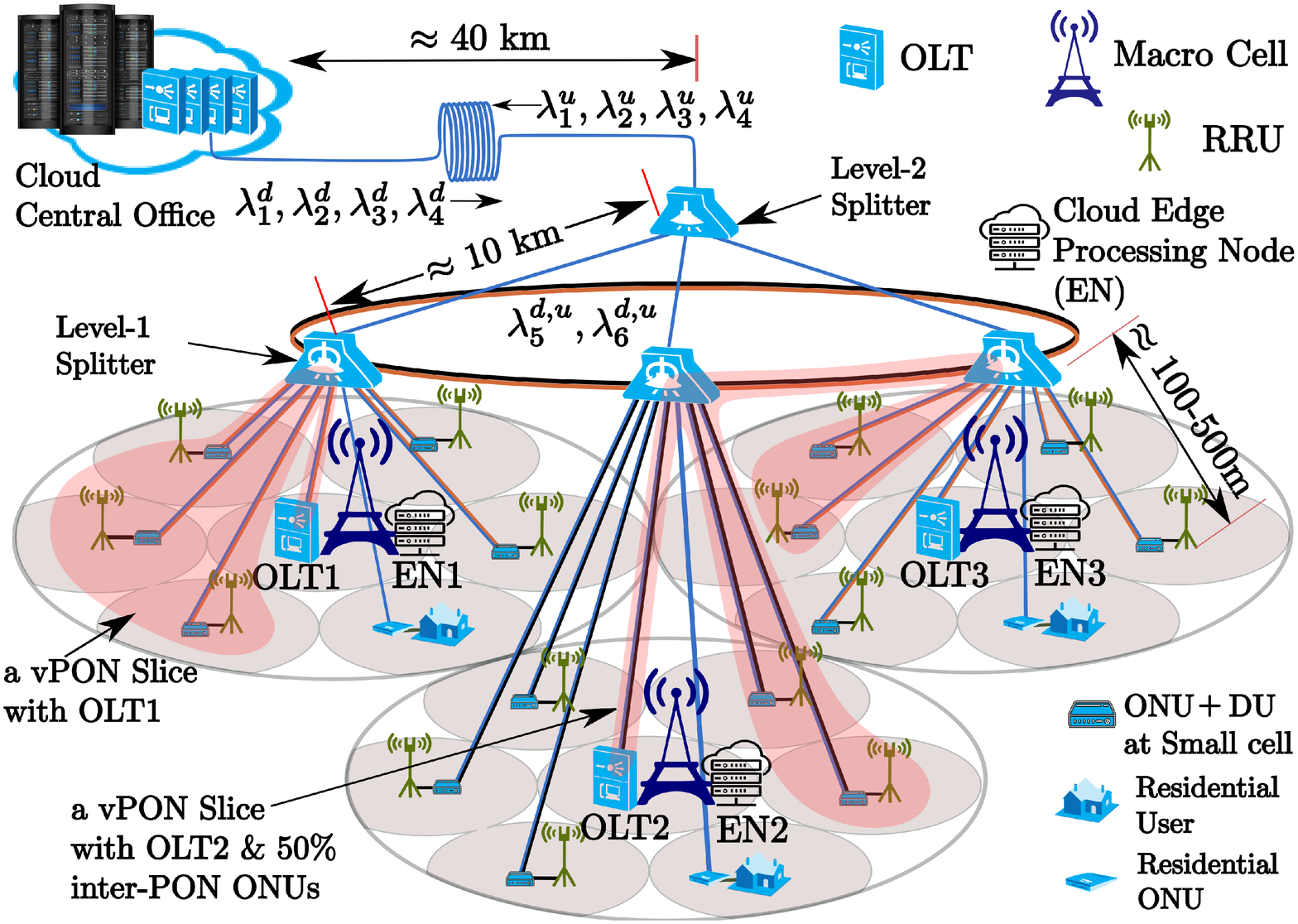}}
        \caption{\small{System Architecture.}}
        \label{Fig:SystemArc}
    \end{subfigure}%
    \,\,
    \begin{subfigure}{0.46\textwidth}
        \frame{\includegraphics[width=\textwidth,right]{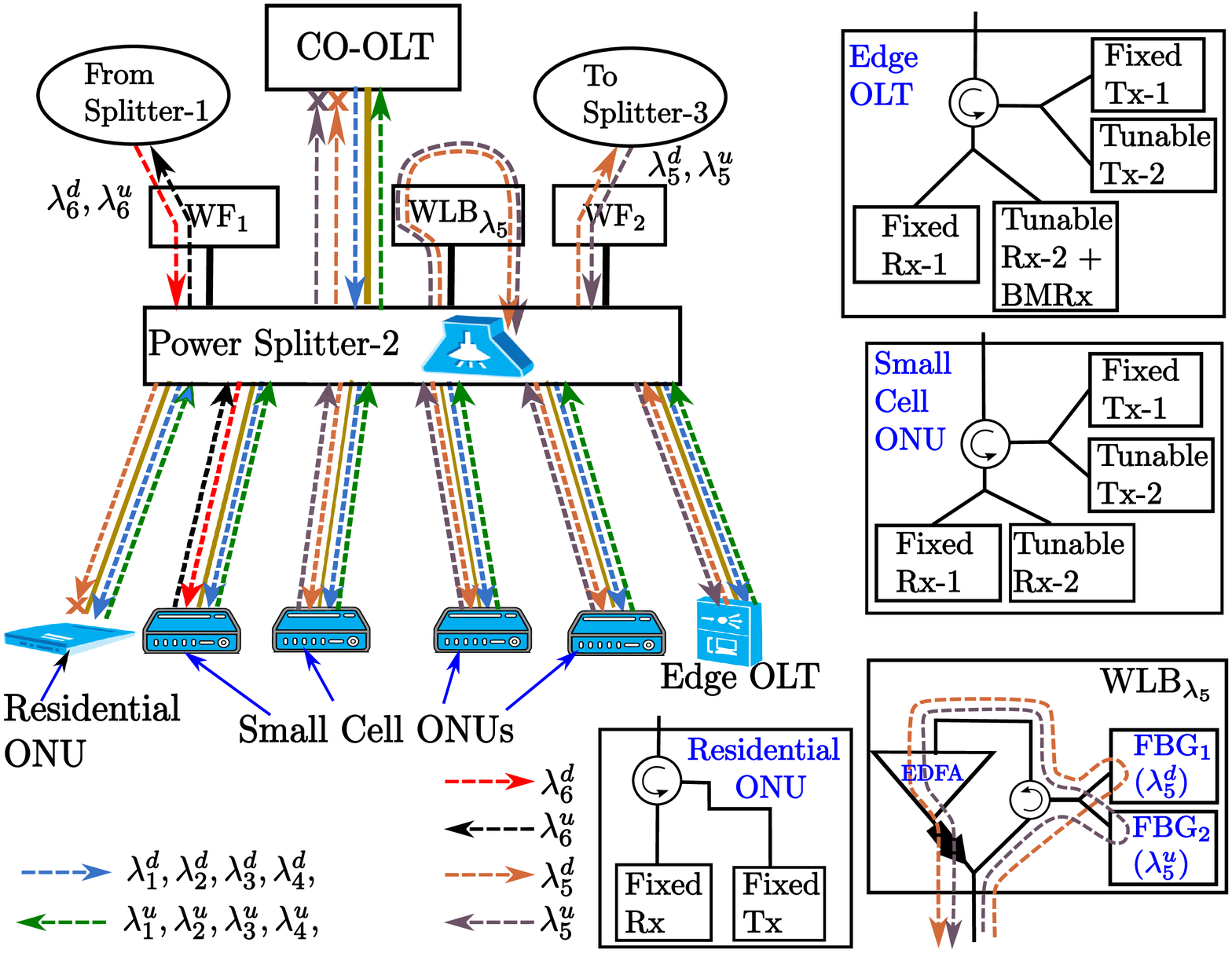}}
        \caption{\small{Architecture of the level-1 splitter.}}
        \label{Fig:SplitterArc}
    \end{subfigure}
    \vspace{-0.25in}
\caption{\small{Architecture of the proposed scheme.}}
\label{Fig:ProposedScheme}
\vspace{-0.25in}
\end{figure}

\vspace{-1.5mm}
Fig. \ref{Fig:ProposedScheme} illustrates the system architecture of our proposed MFH over a \ac{C-RAN} scenario. We consider a \ac{TWDM}-\ac{PON} based mobile fronthaul network, shared with residential users, as shown in the Fig. \ref{Fig:SystemArc}. \acp{RU} are connected with \acp{DU} through a two-stage splitter hierarchy. While our architecture can support multiple scenarios of edge cloud convergence, in this work we consider a popular mobile cell placement strategy, where several small cells are deployed to provide offload capability to a macro cell. 
Further, we consider that \ac{MEC} servers with limited processing capacity are deployed at the macro-cell sites in order to process delay-sensitive traffic. The level-1 splitter connects all the RUs belonging to the coverage area of each macrocell. We refer to this as level-1 \ac{PON} tree. The level-2 splitter interconnects the level-1 splitters to the \ac{CO}. However, unlike \cite{vPON}, we propose an interconnection between level-1 splitters to establish communication between level-1 PON branches. 
It is important to emphasise that this interconnection can be implemented either through direct cable routes between level-1 splitters or else, if existing ducts are not available, as an overlay over the existing level-1 to level-2 splitters' fiber routes (the difference in performance is shown later in Fig. \ref{Fig:LatencyVsNumONUsComp}).
Fig. \ref{Fig:SplitterArc} presents the architecture of the proposed level-1 splitter.

Each level-1 splitter uses three additional blocks namely \ac{WLB}$_{\lambda_i}$, \ac{WPF}$_1$ and \ac{WPF}$_2$, where $\lambda_i$ is the operating wavelength of the edge OLT. Each block connects to the upper side of the splitter, which can have as many ports as there are on the lower side (splitter are inherently symmetrical in this sense). $\text{WLB}_{\lambda_i}$ makes use of two reconfigurable \acp{FBG} connected through a coupler, a circulator and, where required (depending on the splitter loss), a semiconductor optical amplifier, to reflect back selected wavelengths towards the edge. Therefore, if the operating wavelength of the edge OLT is $\lambda_5$, $\text{WLB}_{\lambda_5}$ (as in Fig. \ref{Fig:SplitterArc}), reflects back $\lambda_5^d$ (for downstream) and $\lambda_5^u$ (for upstream) towards the edge OLT. This enables the edge OLT to connect to the ONUs of its own level-1 PON tree (the \ac{vPON} is shown as red shaded area in Fig. \ref{Fig:SystemArc}. 
$\text{WPF}_2$ let $\lambda_5^d$ and $\lambda_5^u$ pass through to connect to the upper side of the splitter of the adjacent level-1 PON tree, enabling the edge OLT to connect to the ONUs of its neighbouring level-1 PON tree (also shown as red shaded area). This enables a \ac{vPON} that has more than one edge OLT to communicate using the same wavelength. Similarly, $\text{WPF}_2$ let the operating wavelength of the edge-OLT of the adjacent level-1 PON tree ($\lambda_6^d$ and $\lambda_6^u$ in this case) pass through, enabling the ONUs of the current Level-1 PON tree to send and receive upstream and downstream traffic to/from the neighbouring edge-OLT. 
\pagestyle{empty}

In addition, if more capacity is required, additional wavelengths (e.g., $\lambda_7$) can be provisioned for the communications to edge OLTs in the adjacent PON branch.
\vspace{-2mm}

The \ac{CO}-\ac{OLT} can employ a one-channel XGS-PON or a TWDM PON with four (or more) wavelengths ($\lambda_1^i, \dots, \lambda_4^i \,\, | \,i \in {d,u}$) for upstream and downstream. $\lambda_1^{d,u}$ is dedicated for exchanging control information such as wavelength reconfiguration and \ac{vPON} slice information in the \ac{MFH} with small cell \acp{ONU} and edge \acp{OLT}. The surplus bandwidth of $\lambda_1$ along with $\lambda_2^i, \dots, \lambda_4^i \,\, | \,i \in {d,u}$ is shared with other users. In order to dynamically connect to the edge \ac{OLT} and \ac{CO}-\ac{OLT}, each small cell ONU employs one fixed (i.e., to reduce cost) and one tunable transceiver, so that  a control channel to the CO is always available.
The edge OLT, hosting the MEC node, also employs a similar pair of transceiver structure with the fixed transceiver dedicated for control message exchange with \ac{CO}-\ac{OLT} and the tunable one used for the datapath of the dynamic \ac{vPON} slices. Other residential \acp{ONU} can adopt single wavelength XGS-PON or else tunable TWDM, depending on their requirements (and suitable cost).
The \ac{vPON} slice allocation is carried out at the CO and communicated to the edge \acp{OLT} through \ac{PLOAM} messages from the \acp{OLT} located at the CO.

\vspace{-1mm}
\section{Performance Evaluation and Results}\vspace{-1mm}
 The topology we consider is that of a converged access/metro architecture \cite{LRPON,super-pon}, where the main \ac{CO} is located 50km away from the edge. Of this, 40 km are used by the main feeder fibre, 10 km by the distance between level-1 and level-2 splitters, while the distance from the last splitter to the edge is up to 500 meters. This is an example of a popular converged access/metro architecture \cite{super-pon}, currently under standardisation, although the proposed system can support different distance distributions.
 We simulate the proposed architecture using the discrete event simulator OMNET++, where each wavelength channel follows the XGS-PON specification. The small cells implement C-RAN with \ac{LLS}, as described in \cite{NGMN}, where each \ac{RU} is served by an ONU and the OLT, \ac{DU}, and \ac{CU} is either at the edge (\ac{MEC}) or \ac{CO}. 
 The core network functions are hosted at the CO regardless of the placement of CU/DU. The traffic from \ac{RU} to \ac{DU} is modeled as \ac{eCPRI} traffic. We consider split-8, operating over a \ac{VRF} scheme \cite{VRF} and split-7.1 \cite{NGMN}, both providing variable rate depending on the actual traffic at the cell. The user arrival process is modeled as a Poisson distribution. The fronthaul rates for split-7.1 are derived from \cite{SplitPhy}: as the cell bandwidth varies from 1.4 to 20MHz, for an RU having two antennas, the fronthaul rate for split-8 goes from 153 to 2,457 Mb/s, while the split-7.1 from 110 to 1,058 Mb/s.
 
 The \ac{DBA} process works as follows. Each \ac{OLT} with a prior scheduling information of UE from the CU (i.e., following the cooperative DBA approach \cite{G.989.3Am1}), schedules the entire \ac{eCPRI} payload output of the \ac{ONU} for the corresponding \ac{TTI} (1ms). Considering a \ac{DBA} cycle  of 125$\mu s$, the \ac{OLT} is required to schedule the entire \ac{eCPRI} payload over 8 grant cycles. As the \ac{RU} requires some local processing time for the \ac{LLS} functional split processing and the encapsulation of the \ac{eCPRI} traffic, we model this with uniform distribution with an upper limit of 125 $\mu$s.
 
 
\begin{figure}[H]
    \quad \,
    \\
    \begin{minipage}{0.312\textwidth}
        \vspace{-0.2in}
        \includegraphics[width=\textwidth,right]{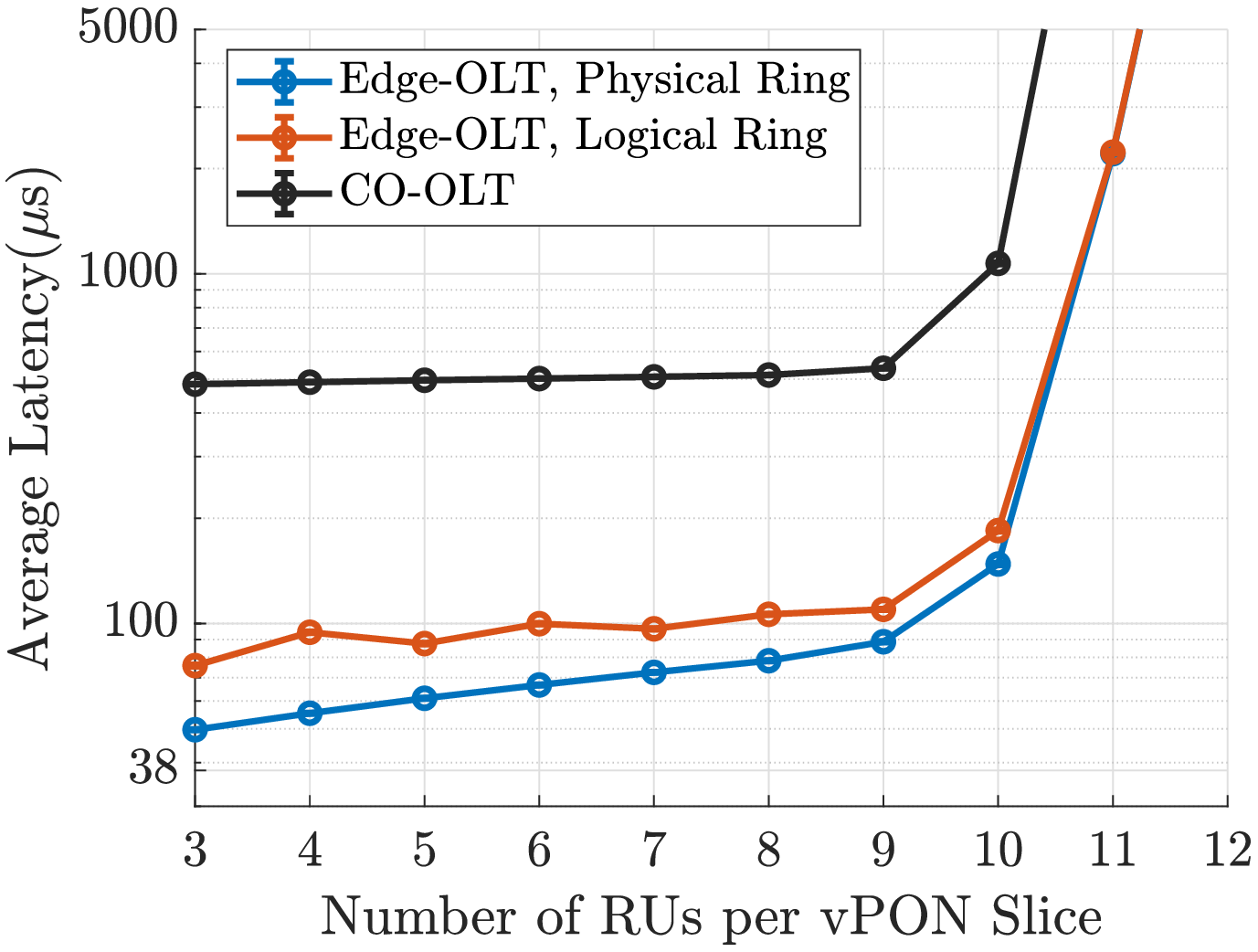}
        \vspace{-0.25in}
        \captionsetup{width=0.9\textwidth}
        \caption{\small{VRF MFH transport latency ($\mu s$) w.r.t vPON slice size for traffic intensity of 12.5 Erlang.}}
        \label{Fig:LatencyVsNumONUsComp}
    \end{minipage}
    \begin{minipage}{0.34\textwidth}
        \vspace{-0.2in}
        \includegraphics[width=\textwidth,right]{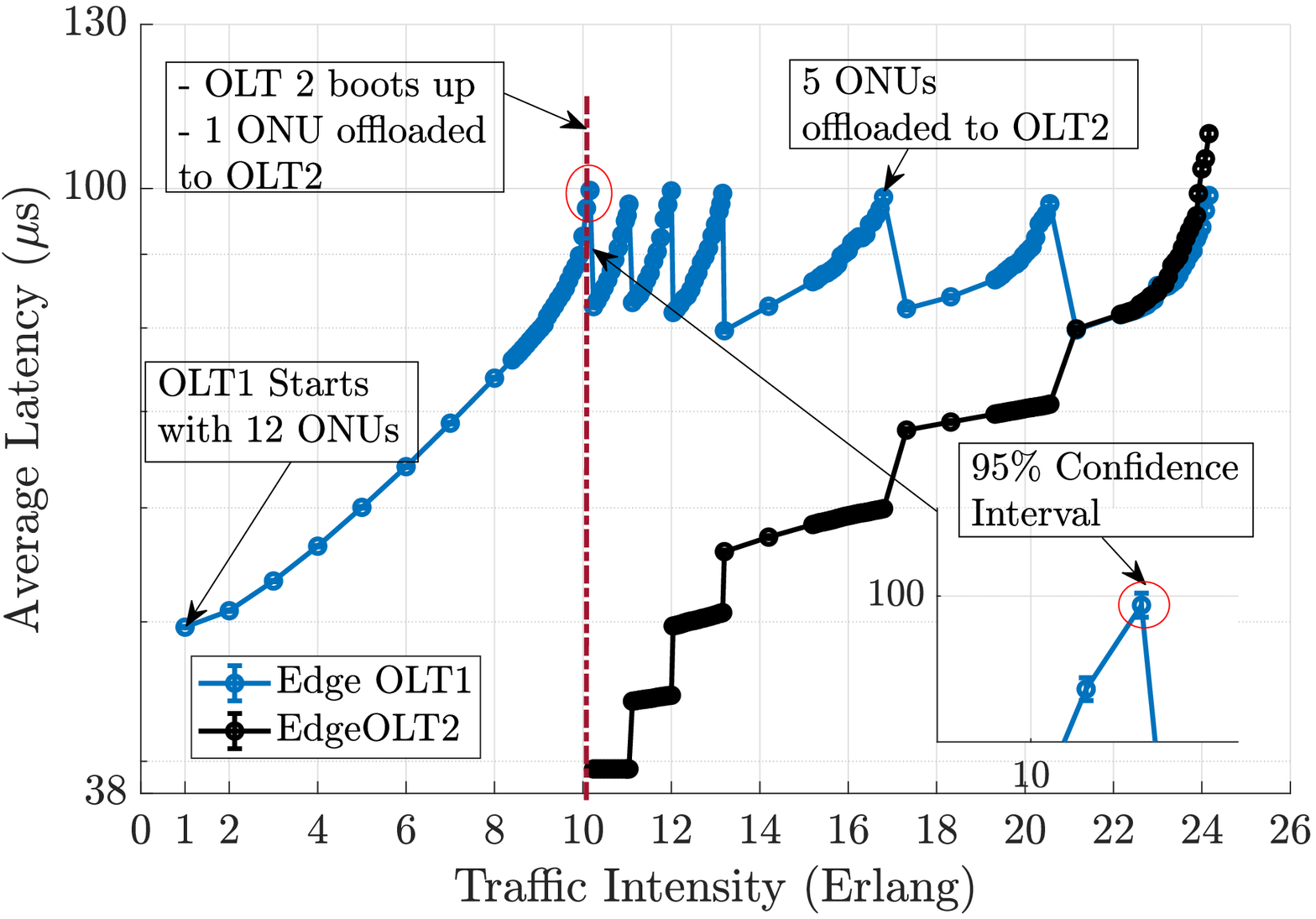}
        \captionsetup{width=0.95\textwidth}
        \vspace{-0.25in}
        \caption{\small{MFH transport Latency w.r.t traffic intensity with unbalanced ONU offloading}}
        \label{Fig:LatencyVsTraffic-EdgeOltDynamicConfigComp}
    \end{minipage}
    \begin{minipage}{0.34\textwidth}
        \vspace{-0.2in}
        \includegraphics[width=\textwidth,right]{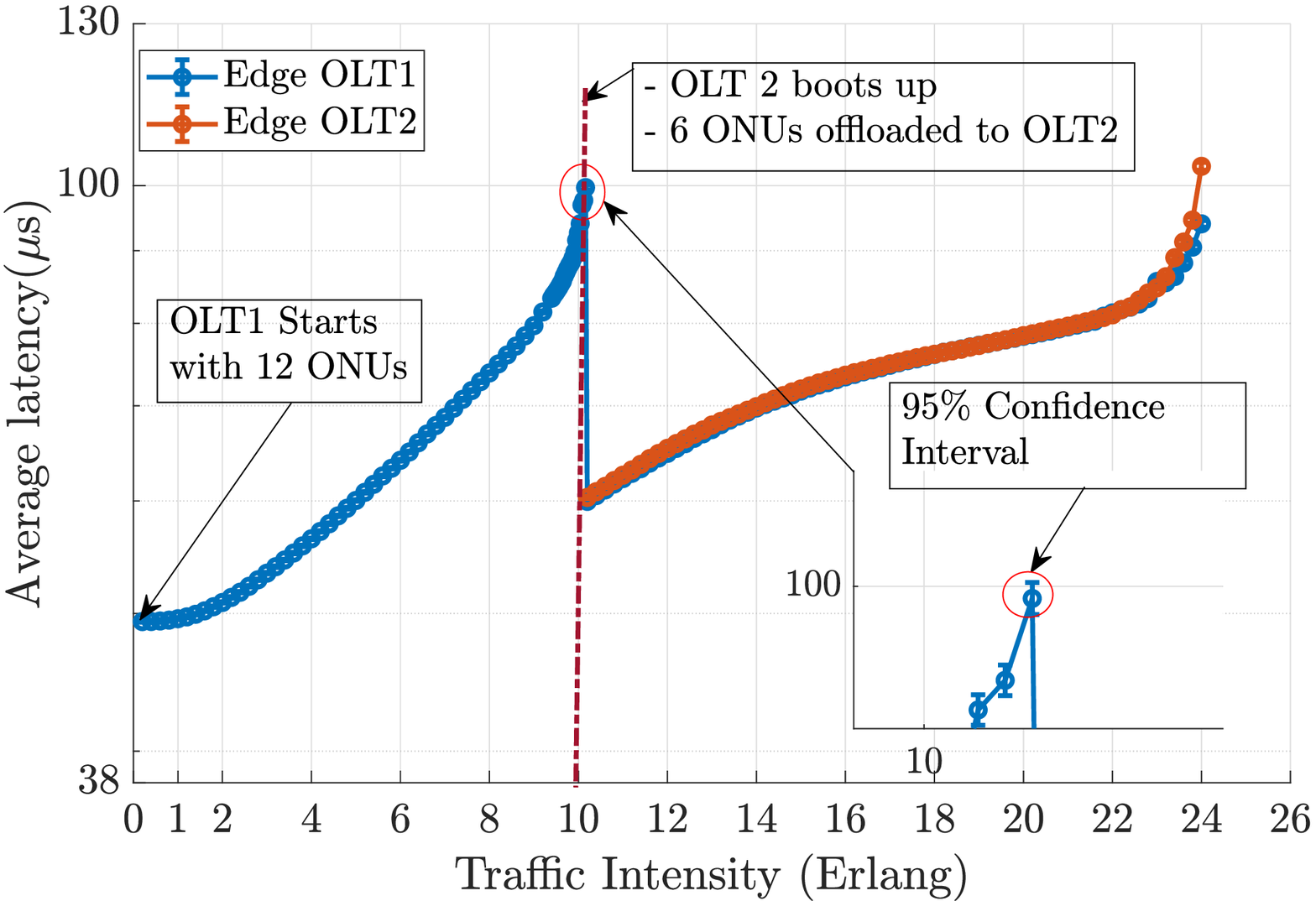}
        \vspace{-0.25in}
        \captionsetup{width=0.95\textwidth}
        \caption{\small{MFH transport Latency w.r.t traffic intensity with balanced ONU offloading}}
        \label{Fig:LatencyVsTraffic-EdgeOltDynamicConfigS2Comp}
    \end{minipage}
    \vspace{-0.2in}
\end{figure}

\vspace{-2mm}
  Fig. \ref{Fig:LatencyVsNumONUsComp} shows a latency reduction of over 10 times between \ac{RU} and \ac{DU}, obtained by edge vPON slicing w.r.t. the use of \acp{OLT} located at the CO. The figure also shows the difference in latency when the fibre routes are overlaid on top of current PON routes (red curve), which is close to 100 $\mu s$, thus compatible with our selected threshold. 
  Fig. \ref{Fig:LatencyVsTraffic-EdgeOltDynamicConfigComp}, illustrates how our proposal can be exploited to considerably improve statistical multiplexing of cells through \ac{MEC} migration of \acp{DU}, by dynamically reconfiguring \ac{vPON} slices, depending on the traffic intensity reports from the \ac{DU}. 
  We consider two edge \acp{OLT} and 24 \acp{ONU}, where half of them are residential and served by the \ac{CO}. Initially, at low traffic volumes, edge OLT1 starts with all the 12 C-RAN \acp{ONU} and we can see that the latency increases as the traffic at each \ac{RU} increases. At 10 Erlang traffic per \ac{RU}, the latency reaches our threshold, set at 100 $\mu s$ (this value can be set to the most appropriate value required by the service). The \ac{CO} thus activates edge OLT2 and reconfigures the \ac{vPON} slices, offloading one \ac{ONU} to the \ac{vPON} slice served by OLT2. This causes a sharp reduction of uplink transport latency at OLT1. The process is then reiterated as soon as the latency grows close to the threshold level. Fig.  \ref{Fig:LatencyVsTraffic-EdgeOltDynamicConfigS2Comp} shows another strategy of balanced-load \ac{ONU} offloading, where 6 of the 12 \acp{ONU} are offloaded simultaneously to the \ac{vPON} slice corresponding to the OLT2, thereby reducing the frequency of offloading events.
  Although not shown here due to space constraints, we have verified experimentally that the backscattering from the WLB action does not affect the system performance, as we achieve a BER of $10^{-5}$ for 10Gb/s transmissions rate even over a 10km (edge) transmission distance.
   
  In conclusion, our mechanism enables using PONs as interconnect for \ac{MEC}-based scenarios, where dynamic PON slicing improves statistical multiplexing of \acp{RU}, while keeping latency below threshold (100 $\mu s$ in our case).
  \vspace{-3mm}
 \section*{Acknowledgement}
\vspace{-2mm}
Financial support from SFI grants 14/IA/252 (O'SHARE) and 13/RC/2077 is gratefully acknowledged.
\vspace{-3mm}


\bibliographystyle{IEEEtran}
\bibliography{IEEEabrv,references}

\begin{thebibliography}{1}
\providecommand{\url}[1]{#1}
\csname url@samestyle\endcsname
\providecommand{\newblock}{\relax}
\providecommand{\bibinfo}[2]{#2}
\providecommand{\BIBentrySTDinterwordspacing}{\spaceskip=0pt\relax}
\providecommand{\BIBentryALTinterwordstretchfactor}{4}
\providecommand{\BIBentryALTinterwordspacing}{\spaceskip=\fontdimen2\font plus
\BIBentryALTinterwordstretchfactor\fontdimen3\font minus
  \fontdimen4\font\relax}
\providecommand{\BIBforeignlanguage}[2]{{%
\expandafter\ifx\csname l@#1\endcsname\relax
\typeout{** WARNING: IEEEtran.bst: No hyphenation pattern has been}%
\typeout{** loaded for the language `#1'. Using the pattern for}%
\typeout{** the default language instead.}%
\else
\language=\csname l@#1\endcsname
\fi
#2}}
\providecommand{\BIBdecl}{\relax}
\BIBdecl

\bibitem{M-DBA}
T.~Tashiro \emph{et~al.}, ``{A novel DBA scheme for TDM-PON based mobile
  fronthaul},'' in \emph{OFC 2014}.

\bibitem{G.989.3Am1}
``40-gigabit-capable passive optical networks: {TC} layer specification amd.
  1,'' ITU-T, Standard, Nov. 2016.

\bibitem{vPON}
R.~I. {Tinini} \emph{et~al.}, ``{Low-latency and energy-efficient BBU placement
  and VPON formation in virtualized cloud-fog RAN},'' \emph{JOCN}, April 2019.

\bibitem{LRPON}
M.~{Ruffini} \emph{et~al.}, ``Access and metro network convergence for flexible
  end-to-end network design,'' \emph{JOCN}, June 2017.

\bibitem{super-pon}
``{IEEE} p802.3cs increased-reach ethernet optical subscriber access
  (super-pon) task force,'' Nov. 2018,
  \url{http://www.ieee802.org/3/minutes/nov18/1118_spon_close_report.pdf}.

\bibitem{NGMN}
N.~Alliance, ``{NGMN} {Overview} on 5g {RAN} {Functional} {Decomposition},''
  Feb. 2018.

\bibitem{VRF}
S.~{Das} \emph{et~al.}, ``{A Variable Rate Fronthaul Scheme for Cloud Radio
  Access Networks},'' \emph{JLT}, July 2019.

\bibitem{SplitPhy}
U.~{Dötsch} \emph{et~al.}, ``Quantitative analysis of split base station
  processing and determination of advantageous architectures for lte,''
  \emph{Bell Labs Technical Journal}, June 2013.

\end{thebibliography}










\end{document}